\documentclass[aps,english,prl,floatfix,superscriptaddress,twocolumn,amsmath,amssymb,longbibliography,nofootinbib]{revtex4-2}

\usepackage[T1]{fontenc}
\usepackage{xcolor}
\usepackage{textcomp}
\usepackage{amsmath}
\usepackage{amssymb}
\usepackage{graphicx}
\usepackage{wasysym}
\usepackage{gensymb}

\makeatletter

\usepackage{bbold}
\usepackage{dsfont}
\PassOptionsToPackage{caption=false}{subfig} 
\usepackage{hyperref}
\hypersetup{
breaklinks=true,
colorlinks=true,
citecolor=blue,
linkcolor=blue,
filecolor=blue,
urlcolor=blue
}
\IfFileExists{lmodern.sty}{\usepackage{lmodern}}{}

\makeatother

\usepackage{babel}
\begin{document}
\title{Asymptotically exact solution of the non-Hermitian disordered interacting Hatano-Nelson chain}
\author{Valéria M. Mattiello}
\affiliation{Gleb Wataghin Physics Institute, University of Campinas (UNICAMP), Rua Sérgio Buarque de Holanda, 777, 13083-859 Campinas, SP, Brazil}
\author{Victor L. Quito \thanks{vquito@ifsc.usp.br}}
\affiliation{S\~{a}o Carlos Institute of Physics, University of S\~{a}o Paulo, IFSC – USP, 13566-590, Sao Carlos, SP, Brazil.}
\author{Eduardo Miranda}
\affiliation{Gleb Wataghin Physics Institute, University of Campinas (UNICAMP), Rua Sérgio Buarque de Holanda, 777, 13083-859 Campinas, SP, Brazil}

\date{\today}
\begin{abstract}
We present an asymptotically exact solution of a paradigmatic non-Hermitian model: the disordered interacting fermionic Hatano-Nelson model, or equivalently, the non-Hermitian spin-1/2 XXZ model. We use a renormalization group method suited for disordered systems and show that non-Hermitian couplings are relevant perturbations to the Hermitian model, which ultimately leads to a quantum-to-classical crossover. The ground state of the model consists of a collection of strongly coupled pairs of spins of arbitrary size at random positions which, unlike the Hermitian case, do not form singlets, but a mixture of the singlet and the $M=0$ triplet state. As a result, 
the magnetic susceptibility in the $x,y$-directions becomes negative and diverges at a finite small temperature. Additionally, in sharp contrast to the $\ln(L)$ increase observed in disordered Hermitian chains, the entanglement entropy of a partition of size $L$ saturates for large $L$, as the strongly coupled pairs become classical and stop contributing at large length scales. 
\end{abstract}
\maketitle

\textit{Introduction} -- 
The study of open quantum systems originated in the pursuit of a description of the effects of various external perturbations, especially those coming from external baths (see, e.g., \cite{Rotter2015}). It has found applications in acoustics~\cite{gu_int_app_2021}, optics~\cite{Zhang_optics_2018}, photonics~\cite{feng_int_app_2017} and other areas. More recently, it has seen a resurgence in situations where the quantum system is submitted to intermittent measurements \cite{Wiseman2009}. In either case, the system is often described by a Lindblad equation \cite{Daley2014,Minganti2019}, part of which relies on a non-Hermitian (nH) Hamiltonian dynamics (for a review, see~\cite{review_Ueda_2020}). 

Although most nH studies have focused on clean systems, some attention has been given to the effects of disorder \cite{hebert_hatano_nelson_2011,Hamazaki2020, Longhi2020, Claes_HN_2021, Zhang2023a,
Kokkinakis2024, Midya2024, Wang2025,Wang2025a,shang2025,sun2025,Longhi2025,Li2025,Kawakami_PRL_2024}. Moreover, the effects of interparticle interactions have mostly been looked at in the absence of disorder \cite{Albertini1996,Bilstein1997,Fukui1998,couvreur_entanglement_2017, Zhang2021a,Chen_HN_2023,Sayyad2023,Li2023,Yang2024, Alcaraz2024, Mao_NH_2024, Gernot_PRR_2025}. Finally, when both disorder and interactions are present, there has been a lot of  numerical work \cite{Liu_localization_2023,Lu2024,Suthar2025,brighi2025}, with little analytical insight (for a notable exception, see \cite{tiwary2025}). In this work, we intend to fill this gap with an asymptotically exact low-energy solution of the disordered, nH antiferromagnetic spin-1/2 XXZ chain, or equivalently, the disordered interacting nH fermionic Hatano-Nelson model~\cite{HatanoNelson1,HatanoNelson2, HatanoNelson3,Dora_HN_2022,orito_entanglement_2023,Chen_HN_2023,Mao_NH_2024}. We will show that: (\textit{i}) it is governed by an  infinite-disorder low-energy fixed point~\cite{Fisher_PRB_Ising_1995}, (\textit{ii}) the fixed-point distributions can be obtained analytically and demonstrate that disorder in the nH couplings is relevant in the renormalization group (RG) sense, leading to a quantum-to-classical crossover, (\textit{iii}) the entanglement entropy of a partition of size $L$ acts as a witness for said crossover, and finally, (\textit{iv}) the spin susceptibility (of the spin model) becomes highly anisotropic at low temperatures, in sharp contrast to the Hermitian case, where all directions show the same power-law behavior. Other known properties of the non-interacting case can also be easily inferred from the solution.

\textit{The model} -- We focus on the disordered nH spin-$1/2$ XXZ chain 
\begin{equation}\label{eq:spin_op_H}
    H = \sum_{i}J_{i}\left[\frac{e^{\gamma_{i}}}{2}S_{i}^{+}S_{i+1}^{-}+\frac{e^{-\gamma_{i}}}{2}S_{i}^{-}S_{i+1}^{+}+\Delta_{i}S_{i}^{z}S_{i+1}^{z}\right].
\end{equation}
The couplings $J_{i},\Delta_{i}$, and $\gamma_{i}$ are spatially uncorrelated and chosen from three independent probability distributions with the restrictions $J_{i}>0$ and $\Delta_{i}>-1/2$ to avoid ferromagnetic phases. Within these assumptions, this represents the most general parametrization of the three independent terms that leads to a real spectrum \cite{MattielloSI2024}, which allows us to work only in the unbroken PT-symmetric phase \cite{bender_PTsymm_real_1998}. Eq.~\eqref{eq:spin_op_H} reduces to the XXZ model in the Hermitian limit $\gamma_{i}=0$. It can also be mapped onto fermions through a Jordan-Wigner transformation,

\begin{equation}\label{eq:J-W-H}
H=\sum_{i}J_{i}\left[\frac{e^{\gamma_{i}}}{2}c_{i}^{\dagger}c_{i+1}+\frac{e^{-\gamma_{i}}}{2}c_{i+1}^{\dagger}c_{i}+\Delta_{i}\delta\hat{n}_{i}\delta\hat{n}_{i+1}\right],
\end{equation}
where $\delta\hat{n}_{i}=\hat{n}_{i}-1/2$. This is the disordered interacting fermionic Hatano-Nelson model \cite{HatanoNelson1,HatanoNelson2,HatanoNelson3}. Note that the ground state of this model corresponds to the half-filled case with $\langle n_i \rangle = 1/2$.
We analyze the model with the help of the strong disorder renormalization group (SDRG)~\cite{madasgupta,madasguptahu}. The SDRG, since its conception, has provided deep insights into many disordered systems (for a review, see~\cite{Igloi_EurophysB_2018}), including the Hermitian version of Eq.~\eqref{eq:J-W-H}~\cite{fisher94-xxz}. It has been recently used in a nH setting \cite{tiwary2025}.
The procedure amounts to rescaling states and physical quantities to lower energies and longer wavelengths. 
In the Hermitian XXZ chain (and, as we will show, also in the nH one)
this leads asymptotically to infinitely large effective disorder and, thus, to exact statements about the behavior of the system.
In this paper, we employ it to solve the model above.

\textit{Method and procedure} -- 
The SDRG procedure works by considering the hierarchy of local excitation gaps obtained from solving all the problems of two adjacent spins. We focus on the most strongly coupled pair, the one with the largest gap (which we call $\Omega$) between the ground and first excited state, keep only the ground state, and discard the excited states. We will refer to these pairs of spins as strongly coupled pairs (SCP). This approach is justifiable as long as the energy or temperature scale $T$ of interest and the neighboring couplings are significantly smaller than $\Omega$. It can be shown~\cite{MattielloSI2024} that the local two-spin gap is $J_i(1+\Delta_i)/2$, if $\Delta_i\in [-1/2,1]$, and $J_i$ if $\Delta_i>1$. Because of non-Hermiticity, right and left eigenstates (i.e., eigenstates of $H$ and $H^{\dagger}$) are different. Irrespective of the couplings, the local right ground state is always non-degenerate and written as $\left|GS^{\left(R\right)}\right\rangle =\cosh\frac{\gamma}{2}\left|0,0\right\rangle +\sinh\frac{\gamma}{2}\left|1,0\right\rangle $, where $\left|J,M\right\rangle$ denote the standard total angular momentum states of the pair. The left ground state $\left|GS^{\left(L\right)}\right\rangle$ is obtained by changing $\gamma\rightarrow-\gamma$.
Note that they are a mixture of the singlet and the $M=0$ triplet states, but reduce to pure singlets only in the Hermitian case ($\gamma_i=0$).

\begin{figure}
    \centering
    \includegraphics[width=0.95\linewidth]{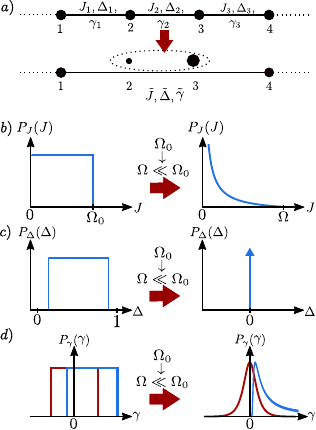}
    \caption{ (a) The SDRG decimation step. The strongest-coupled pair is assumed to be formed by spins 2 and 3. This pair is decimated, and quantum fluctuations lead to effective couplings between spins 1 and 4, initially not directly coupled. The flows of the distributions of the couplings  (b) $J$,  (c) $\Delta$, and (d) $\gamma$, at two different stages of the RG flow. 
    The fixed point of the $J$ distribution is universal and tends towards infinite width. If the anisotropy distribution, $P_{\Delta}(\Delta)$, has support between zero and one, it flows to a delta-function fixed-point distribution at the origin. The distribution of $\gamma$ is shown for both symmetric (red) and asymmetric (blue) initial distributions. Under the flow, both distribution widths increase without limit, but the asymmetric one has support only for $\gamma>0$.}
    \label{fig:distributions}
\end{figure}

After freezing the SCP in its ground state, we treat its coupling to the adjacent spins perturbatively. 
Importantly, perturbation theory has to be implemented using the biorthogonal basis~\cite{Brody_2014} so that the right and left eigenstates are orthonormalized as $\langle \psi^L_i | \psi^R_{j} \rangle = \delta_{i,j}$~\cite{MattielloSI2024}. Assuming that the largest local gap $\Omega$ is between sites $i$ and $i+1$, new renormalized couplings appear between spins $i-1$ and $i+2$ given by~\cite{MattielloSI2024} [see Fig.~\ref{fig:distributions}(a)]
\begin{equation}\label{fisher_c}
    \Tilde{J}_{i-1} = \frac{J_{i-1} J_{i+1}}{\Omega} \frac{1}{1+\Delta_i}, \hspace{0.3cm} \Tilde{\Delta}_{i-1} = \frac{\Delta_{i-1} \Delta_{i+1}(1+\Delta_{i})}{2},
\end{equation}
\begin{equation}\label{non-herm_c}
    \Tilde{\gamma}_{i-1} = \gamma_{i-1} + \gamma_i + \gamma_{i+1}, \hspace{0.3cm}
    \Tilde{\ell}_{i-1} = \ell_{i-1} + \ell_i + \ell_{i+1}.
\end{equation}
We have added, for later convenience, the decimation rule of the bond size $\ell_i$, which is geometrically obvious and identical to that for $\gamma_i$. Note the decoupling of the decimation rule of $\gamma_i$ from those of $J_i$ and $\Delta_i$. This facilitates greatly the analysis of the SDRG flow and justifies the parametrization chosen in Eq.~\eqref{eq:spin_op_H}. The procedure is iterated until we reach either the ground state or the target temperature $T = \Omega \ll \Omega_0$, where $\Omega_0$ is the largest local gap of the bare distributions.

From the structure of the SDRG iterations, we can already conclude what the structure of the ground state is. It will consist of nested SCPs of all sizes formed from spins at random positions. We refer to this phase as a random strongly coupled pair phase (RSCPP). This nomenclature sets it apart from the random singlet phase of the Hermitian model, the XXZ chain~\cite{fisher94-xxz}.

\textit{SDRG flows} - We now analyze the SDRG flow and construct the phase diagram. Due to the decoupling of the $\gamma_i$ variables, the flow of the $J_i$ and $\Delta_i$ distributions \textit{is identical to that of the Hermitian XXZ model}, which we now summarize~\cite{fisher94-xxz}. The relative width of the $J$ distribution increases without limit as the cutoff $\Omega$ is lowered, flowing towards a universal infinite-randomness fixed point (IRFP) of the form
\begin{equation}
\label{IRFP}
    P_J(J) \sim \frac{1}{J^{1-1/\Gamma}}, \qquad \Gamma = \ln\left(\frac{\Omega_0}{\Omega}\right)\to \infty.
\end{equation}
The increasing width of the distribution guarantees that the perturbation theory of Eq.~\eqref{fisher_c} is increasingly more accurate and asymptotically exact at low energies. The flow of the $J$ distribution is schematically shown in Fig.~\ref{fig:distributions}(b).

The infrared stable fixed points of the $\Delta$ distribution are $\Delta_{i}=0$, corresponding to the XX model, and $\Delta_{i}\rightarrow\infty$, corresponding to the Ising fixed point. Separating them is the unstable fixed point of the Heisenberg model $\Delta_{i}=1$~\cite{fisher94-xxz}. For concreteness, we show schematically the first kind of flow in Fig.~\ref{fig:distributions}(c). From now on, we will focus only on flows to the XX and Heisenberg fixed points.

The effects of non-Hermiticity are contained in the flow of $\gamma_i$. The structure of Eq.~\eqref{non-herm_c} might suggest that the $\gamma_i$ would follow a standard random walk. This is not the case, however, because the renormalized $\gamma_i$ become strongly correlated with $J_i$, as the latter primarily define the  local gaps $\Omega_i$ and thus, which bond will be decimated next. Therefore, as very weak bonds are generated after many RG steps, see Eq.~\eqref{IRFP}, their corresponding $\gamma_i$ will be accordingly very large. The asymptotic form of the joint distribution of $J_i$ and bond lengths $\ell_i$ was obtained analytically for the Hermitian XXZ chain \cite{fisher94-xxz}. That result can be used to immediately determine the joint distribution of $J_i$ and $\gamma_i$, due their identical decimation rules. For later calculations, however, we will also need the joint distributions of $J_i$, $\gamma_i$, and $\ell_i$, which is given in \cite{MattielloSI2024}. 

Let us describe briefly the qualitative flow of $\gamma_i$.
From Eq.~\eqref{non-herm_c}, one can easily see that the Hermitian XXZ chain with $\gamma_{i}=0$ is a fixed point. Also from the additive form of Eq.~\eqref{non-herm_c}, it is simple to see that this fixed point is unstable: for any small initial randomness in $\gamma_i$, the distribution will broaden. Some features of this initial distribution crucially influence the flow. If the initial distribution is an even function of $\gamma$, it remains so under the SDRG, but its width increases without limit. More precisely, the fixed-point distribution is a universal even function of the scaled variable  $y = C_{\gamma}\gamma/ \Gamma$, where $\Gamma \to \infty$, see Eq.~\eqref{IRFP}, and $C_{\gamma}$ is a non-universal constant. If the initial distribution is not an even function of $\gamma$, this initial asymmetry grows with the flow. The asymptotic distribution also becomes increasingly broader, \textit{but its support is confined to either positive or negative values of $\gamma$}. More precisely, if the first moment of the distribution is positive (negative), the fixed-point distribution is a universal function of the scaled variable $y = C_{\gamma} \gamma / \Gamma^{2}$ with strict support for positive (negative) values of $y$. This asymmetric case applies to the flow of bond lengths $\ell_i$, which remain strictly positive~\cite{fisher94-xxz}. The schematic evolution of the $\gamma$ distribution for both cases is illustrated in Fig.~\ref{fig:distributions}(d).
This extremely asymmetric form of the $\gamma$ distribution at low energies translates into a dominant one-sidedness of the hopping amplitudes in Eq.~\eqref{eq:J-W-H}. This provides a natural explanation for the non-Hermitian skin effect of the model~\cite{Longhi2020, Claes_HN_2021,Zhang2023a,Midya2024}, \textit{even in the interacting case}, and its complete absence in the symmetric case~\cite{Longhi2025}.

The ground state of the RSCPP is formed out of localized SCPs with arbitrarily long sizes, a complete parallel to the Hermitian case~\cite{fisher94-xxz}. This is a property of the flow and is not spoiled by non-Hermiticity. The localization of the SCPs has a sub-exponential character, however, as observed numerically~\cite{Longhi2025}. This can be seen from the \textit{typical} spin-spin correlation function $\langle S^z_iS^z_j\rangle_{\mathrm{typ}} \sim \exp{-\sqrt{|i-j|/\xi}}$~\cite{fisher94-xxz,Getelina2020}. This was also known for the non-interacting Hermitian fermionic model~\cite{Inui1994}.
 
\begin{figure}
    \centering
    \includegraphics[width = \linewidth]{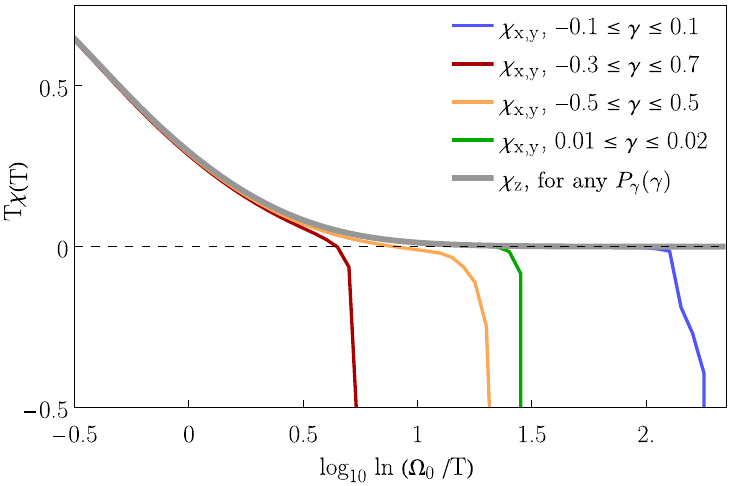}
    \caption{Magnetic susceptibility multiplied by the temperature for different initial uniform distributions of $\gamma$, as a function of $\ln(\Omega_{0}/T)$, in log scale. The $z$-component, $T\chi_z(T)$, shown in gray, is independent of $P_\gamma(\gamma)$ and the same as for the Hermitian chain, approaching zero as $T\to 0$. In red, yellow, green, and blue are $T\chi_{x,y}(T)$, which depend strongly on the distribution of $\gamma$. For small $T$, it becomes negative and diverges at a small by finite $T$, a consequence of a negative contribution from each strongly coupled pair.}
    \label{fig:suscep}
\end{figure} 

\textit{Thermodynamic properties} - We focus first on the magnetic susceptibility of the spin model along the $x,y$ and $z$ directions. 
There are in general two contributions to the magnetic susceptibility $\chi$ at temperature $T$,
\begin{equation}
\chi^\alpha(T)=\chi^\alpha_{\text{free}}(T)+\chi^\alpha_{\text{SCP}}(T), \quad \alpha=x,y,z
\end{equation}
the first term coming from spins not yet decimated and the second one from the already decimated SCPs at scale $\Omega=T$. Because the $J$ distribution is so broad at low-$T$, non-decimated spins are only weakly coupled to their active neighbors and can be accurately treated as free spins. Their contribution is Curie-like, $\sim 1/T$, times the density of active spins at $T$, $n_{\mathrm{free}}(T)\sim 1/[\ln(1/T)]^2$, so that ~\cite{fisher94-xxz}
\begin{equation}
    [\chi^\alpha_{\text{free}}(T)]^{-1}\sim {T[\ln(1/T)]^2}.
\end{equation} 
This contribution is the same for any component of the susceptibility, even if the SU(2) symmetry is broken, $\Delta\ne1$. This is a consequence of an emergent symmetry of the model~\cite{fisher94-xxz,Quito_PRL_2015}. 
The  contribution of a \textit{single} SCP, $\chi_{\mathrm{sscp}}^{\alpha}(T)$, is given by \cite{MattielloSI2024}
\begin{equation}\label{eq:singleSCP}
T\chi_{\mathrm{sscp}}^{\alpha}(T)=
\begin{cases}
-\sinh^{2}\left(\frac{\gamma}{2}\right), & \ \text{if } \alpha = x,y \\
0, & \ \text{if } \alpha = z.
\end{cases}
\end{equation} 
Note how this contribution only exists in the $x,y$-directions and in the nH case. This is because the ground state of a SCP is a superposition of the two-spin singlet and $M=0$ triplet states.
Thus, $\chi_{\text{SCP}}(T)$ involves summing over all the decimated SCPs, weighted by the $\gamma$ distribution at the scale at which they were decimated. A key feature is that $\chi^{{x,y}}_{\text{SCP}}(T)$ \textit{diverges at a finite} $T$. For the symmetric case, for instance,~\cite{MattielloSI2024}
\begin{equation}
\chi^{{x,y}}_{\text{SCP}}\left(T\right)\sim-\frac{1}{4T}\ln\left[\frac{2C_{\gamma}}{\Gamma_{T}}\tan\left(\frac{\Gamma_{T}}{2C_{\gamma}}\right)\right],
\end{equation}
where $\Gamma_{T}=\ln(\Omega_{0}/T)$, which diverges at $T=\Omega_{0}\exp\left(-\pi C_{\gamma}\right)$.
There is also a divergence in the asymmetric case~\cite{MattielloSI2024}. 
This behavior can be traced back to the solution of the nH two-site problem, see Eq.~\eqref{eq:singleSCP}, and is not a consequence of any collective phase transition. 
The susceptibility in the $z$-direction is the same in both Hermitian and nH cases
and remains positive and non-divergent, since no SCP contributes. Fig.~\ref{fig:suscep} shows the total susceptibility 
for a numerical implementation of the SDRG procedure in chains with periodic boundary conditions, in several different cases. It is worth mentioning that the susceptibility \textit{is not a self-averaging quantity}, showing extremely large sample-to-sample fluctuations, which is discussed in more detail in~\cite{MattielloSI2024}.

Finally, at a low temperature $T$, the specific entropy receives no contributions  from the frozen SCPs, and only a density $n_{\mathrm{free}}(T)$ of non-decimated spins contributes $k_B \ln 2$. From this, one obtains the low-temperature specific heat as $c_V(T)\sim k_B/[\ln(1/T)]^3$, which is the same as in the Hermitian case~\cite{fisher94-xxz}.

\begin{figure}
    \centering
    \includegraphics[width=\linewidth]{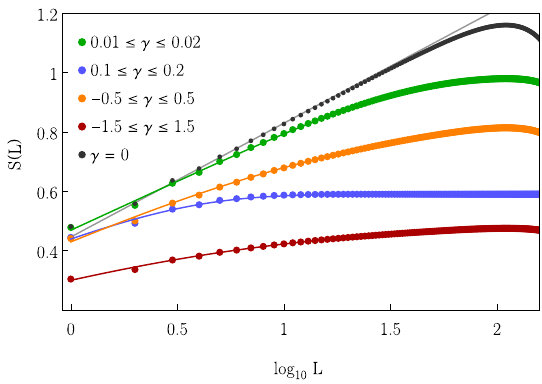}
\caption{Entanglement entropy as a function of the partition size $L$. The points were obtained by numerically decimating chains of length $N=220$ and averaging over 2,000 disorder realizations. The lines correspond to fits to the analytical expressions derived in the asymptotic large-$L$ regime. The RSCPP shows a saturation at large $L$, a consequence of the SCPs formed by spins far apart not contributing as they become classical. In black is the entanglement entropy of the Hermitian model, with the gray line showing the analytical result. In this case, each pair of spins is maximally entangled, which leads to a logarithmic increase of $S(L)$. In all cases, there is a downturn as $L\to N/2$, because $S(L)$ is symmetric with respect to the chain half size.}
    \label{fig:EE}
\end{figure}

\textit{Entanglement entropy} - Another remarkable effect of non-Hermiticity is captured by the ground state entanglement entropy (EE) of a partition of size $L$~\cite{Herviou_entanglement_2019,Chen_2024_CPS}. It is defined by
\begin{equation}
    S(L)=-\text{Tr}\left[\hat{\rho}(L)\log_{2}\hat{\rho}(L)\right] \label{eq:S_pair},
\end{equation}
where $\hat{\rho}(L)$ is the reduced density matrix of the partition, fixing the full system in its ground state. Within the SDRG, 
only spins inside the partition that form SCPs with spins outside of it contribute to the EE. We thus need to add the contributions from those SCPs that cross the partition boundaries ~\cite{RefaelMoore_PRL_2004,Hoyos2007PRB,Refael_2009}. 

There are two possible ways of defining the density matrix of a nH system: $\hat{\rho}^{RL}=\left|\psi^{R}\right\rangle \left\langle \psi^{L}\right|$~\cite{fossati_entanglement_2023,Cipolloni_entanglement_2023,couvreur_entanglement_2017,guo_entanglement_2021,Hsieh_entanglement_2023} or $\hat{\rho}^{RR}=\left|\psi^{R}\right\rangle \left\langle \psi^{R}\right|$~\cite{Herviou_entanglement_2019,modak_entanglement_2021,kawabata_entanglement_2023}
(the case of LL, with left eigenvectors, is identical to RR). Interestingly, these two schemes lead to remarkably different results for the EE of a single SCP, $S_{\mathrm{pair}}$. For the RL case, there is always maximal entanglement in an SCP, $S_{\text{pair}}=1$, and the EE does not depend on the nH coupling $\gamma$~\cite{MattielloSI2024}.
In the RR case, in constrast, the EE depends on $\gamma$~\cite{MattielloSI2024}, 
\begin{equation}
S_{\text{pair}}=\left|\gamma\right|\left(1-\tanh\left|\gamma\right|\right)\log_{2}e+\log_{2}\left(1+e^{-2\left|\gamma\right|}\right).
\label{eq:EE_pair}
\end{equation}

It is equal to 1 only when $\gamma=0$, recovering the behavior of singlets, but decreases to zero as $\left|\gamma\right|\rightarrow\infty$, as the SCP becomes separable. As the distribution width of $\gamma$ grows without limit, long, low-energy SCPs stop contributing to the EE, which saturates for large enough $L$. 

The EE of a size-$L$ partition, $S(L)$, is shown in Fig.~\ref{fig:EE} for different distributions of $\gamma$, obtained by numerical simulations of chains of size $N=220$, using periodic boundary conditions  (circles). Note that $S(L)$ is symmetric with respect to the chain half size $N/2$. Nevertheless, it is clear, especially for broader initial $\gamma$ distributions, that the EE saturates before $L$ reaches $N/2$. This is confirmed by analytical calculations, see the lines in Fig.~\ref{fig:EE}. The calculation requires the use of the fixed-point joint distribution of couplings, bond lengths and $\gamma$, which we have explicitly obtained~\cite{MattielloSI2024}.

\textit{Conclusions} - While a lot of recent work on disordered nH systems has been performed numerically, we have obtained an asymptotically exact analytical description of a disordered nH spin chain, equivalent to the fermionic interacting Hatano-Nelson model. 
We have mapped out its phase diagram and described some of its physical properties. The main result is that the non-Hermiticity is \textit{relevant} in the RG sense, i.e., it grows both in strength and distribution width at low energies. In our case, this is manifested explicitly as a divergence in the magnetic susceptibility at a finite small temperature (in contrast to a quasi-Curie-like behavior in the Hermitian case) and a saturation of the entanglement entropy as a  function of partition size (in contrast to a log-increase in the Hermitian case). Our work opens up the possibility of exploring several other cases, such as nH versions of random transverse-field Ising model. 

\textit{Acknowledgments} - We acknowledge valuable discussions with Francisco Alcaraz and Jed Pixley. VMM acknowledges financial support from Capes grant 88887.687908/2022-00 and CNPq grant 141548/2023-1. VLQ acknowledges financial support from the CNPq grant 311565/2023-9, the University of São Paulo startup grant number 22.1.09345.01.2, and the Fapesp Grant no. 2024/09202-0. EM acknowledges financial support from CNPq (Grant no. 309584/2021-3) and Fapesp (Grant no. 2022/15453-0). 

\bibliographystyle{apsrev4-2}
\bibliography{arxiv_PRR_final}

\newpage

\pagebreak

\section*{Supplemental material}

This Supplemental Material is organized as follows. In Section~\ref{sec:param_H}, we present other possible parametrizations of the Hamiltonian. In Section~\ref{sec:twosite_RGrules}, we present the spectrum of the 2-site problem and the SDRG decimation rule of the couplings. Section~\ref{sec:Distr} is devoted to the analysis of the joint distribution of $J$, $\gamma$ and $\ell$. Finally, we present some technical details of the derivation of two observables: the magnetic susceptibility, in Section~\ref{sec:magn_suscep}, and the entanglement entropy, in Section~\ref{sec:EE}.

\section{Other parametrizations of the Hamiltonian \label{sec:param_H}}

We now analyze the generality of the Hamiltonian of Eq.~(1) of the main text. The most general nearest-neighbor Hamiltonian invariant under $\mathrm{SO}(2)$ rotations around de $z$-axis, which excludes terms such as $S_{i}^{+}S_{i+1}^{+}$ or $S_{i}^{-}S_{i+1}^{-}$, is
\begin{equation}
    H_G = \sum_i \left[A_i S^+_i S^-_{i+1} + B_i S^-_iS^+_{i+1} + C_i S^z_i S^z_{i+1}\right], \label{eq:H_gen}
\end{equation}
where $A_i, B_i$, and $C_i$ are assumed to be uncorrelated and drawn from independent probability distributions. By solving the two-spin problem, we find the following eigenvalues
\begin{equation}\label{eq:eigen_general}
\begin{split}
    &\lambda^G_1 = -\sqrt{A_iB_i} - \frac{C_i}{4}, \hspace{1cm} \lambda^G_3=\lambda^G_4 = \frac{C_i}{4},\\
    &\lambda^G_2 = \sqrt{A_iB_i} - \frac{C_i}{4}.
\end{split}
\end{equation}
The case $A_i=B_i$ corresponds to the Hermitian XXZ model, while $A_{i}=B_{i}=C_{i}$ is the SU(2)-symmetric Hermitian Heisenberg model. From Eq.~\eqref{eq:eigen_general}, we see that the eigenvalues can be complex if $A_i$ and $B_i$ have opposite signs. Therefore, if we insist on working in the PT-symmetric phase, we should assume they have the same sign. Without loss of generality, we take them to be both positive. In this case, the following reparametrization 
\begin{equation}
\begin{split}
    &\gamma_i = \ln{\sqrt{\frac{A_i}{B_i}}},\hspace{1cm} \Delta_i = \frac{C_i}{2\sqrt{A_i B_i}},\\
    &J_i = 2\sqrt{A_i B_i}
\end{split}
\end{equation}
takes Eq.~\eqref{eq:H_gen} into our Hamiltonian of Eq.~(1) of the main text.

This parametrization \cite{Midya2024} is advantageous also because: (1) the eigenvalues become independent of $\gamma_{i}$ [see later Eq.~\eqref{eq:eigenvalues}], (2) the SDRG decimation rule for $\gamma_i$ decouples from the other couplings [see Eqs.~(3) and~(4) of the main text], and (3) the fixed point analysis is significantly simpler.

\section{The two-site problem and the derivation of the SDRG rules \label{sec:twosite_RGrules}}

\subsection{Solution of the two-site problem}
The building block of the SDRG scheme involves solving the two-spin Hamiltonian, here assumed to connect sites $i$ and $i+1$,
\begin{equation}\label{eq:two-site}
    H_{i} =  J_{i} \Bigg[\frac{e^{\gamma_{i}}}{2} S^+_{i}S^-_{i+1} + \frac{e^{-\gamma_{i}}}{2} S^-_{i}S^+_{i+1} + \Delta_{i} S^z_{i} S^z_{i+1}\Bigg].
\end{equation}
The eigenvalues of $H_i$ are
\begin{equation}\label{eq:eigenvalues}
\begin{split}
    &\lambda_{1}=-\frac{J_{i}}{2}\left(1+\frac{\Delta_{i}}{2}\right), \hspace{1cm}    \lambda_{3}=\lambda_{4}=J_{i}\frac{\Delta_{i}}{4}, \\
    &\lambda_{2}=\frac{J_{i}}{2}\left(1-\frac{\Delta_{i}}{2}\right).
\end{split}
\end{equation}
Note that the eigenvalues are independent of $\gamma_i$ and, therefore, the same as those of the Hermitian XXZ model. Provided  $J_i$ and $\Delta_i$ are real numbers, the eigenvalues are all real and the problem is PT-symmetric.  The spectral gaps are 
\begin{equation}
\begin{split}
    &G_{2 \rightarrow 1} = \lambda_2 - \lambda_1 = J_{i},\\
    &G_{3|4 \rightarrow 1} =\lambda_3- \lambda_1 = \lambda_4 - \lambda_1 = \frac{J_{i}}{2} (1 + \Delta_{i}),\\
    &G_{3|4
 \rightarrow 2} =\lambda_3- \lambda_2=\lambda_4 - \lambda_2 = - \frac{J_{i}}{2}(1 - \Delta_{i}).
\end{split}    
\end{equation}
The index $3|4$ is used because $\lambda_3 = \lambda_4$. The right eigenvectors (R) of $H_{i}$ are
\begin{equation}
\begin{split}
    &|\psi^R_1\rangle=\cosh\frac{\gamma_{i}}{2}\left|0,0\right\rangle +\sinh\frac{\gamma_{i}}{2}\left|1,0\right\rangle,\\ 
    &|\psi^R_2\rangle =\sinh\frac{\gamma_{i}}{2}\left|0,0\right\rangle +\cosh\frac{\gamma_{i}}{2}\left|1,0\right\rangle, \\&|\psi^R_3\rangle = |1,1\rangle,\\&|\psi^R_4\rangle = |1,-1 \rangle. \label{eq:states_2spins}
\end{split}    
\end{equation}
These states are written in the basis of total angular momentum of the two sites $|J,M\rangle$, where $J=0,1$ and the $M$-values are obvious. Unlike the eigenvalues, the eigenvectors do depend on the coupling $\gamma_i$ and, therefore, differ from their Hermitian counterparts. The eigenvectors of the Hermitian conjugate of the Hamiltonian, $H_{i}^\dagger$, are the left eigenvectors (L) and are evidently obtained from $|\psi^R_i\rangle$ by making the substitution $\gamma_{i} \to -\gamma_{i}$. This set of eigenvectors is already orthonormalized in the biorthogonal scheme $\langle\psi^L_i|\psi^R_{j} \rangle = \delta_{i,j}$. Note that the right (or left) eigenvectors by themselves may be neither mutually orthogonal nor complete~\cite{Brody_2014}. 

\subsection{Derivation of the SDRG rules}

The first step of the SDRG procedure is to find the pair of neighboring spins that has the largest local gap. We will be working with distributions of positive $J_{i}$ and $\Delta_{i}>-1/2$, thus avoiding possible ferromagnetic phases~\cite{fisher94-xxz}. In this regime, the local two-site ground state has always energy $\lambda_{1}$. In the low-temperature regime, the excited states of this pair will not be thermally accessed, and, therefore, this SCP will remain frozen in its ground state. Quantum fluctuations, however, lead to effective couplings between the spins initially connected to the SCP~\cite{madasgupta,madasguptahu}. The goal is to calculate the value of these effective couplings. Let us assume that the SCP is formed by spins at sites 2 and 3. We then split the Hamiltonian as
\begin{equation}
H=H_{2}+\sum_{i\ne2}H_{i},
\end{equation}
with $H_{i}$ defined in Eq.~\eqref{eq:two-site}. $H_2$ is the unperturbed Hamiltonian. Spins that are not directly connected to the pair are only modified by effects that appear in much higher order in perturbation theory and can, therefore, be neglected. The perturbation to be considered is, therefore,
\begin{equation}
V=H_{1}+H_{3}.
\end{equation}

Perturbation theory can be easily adapted to nH Hamiltonians. The only important modification is the use of the biorthogonal basis~\cite{Brody_2014, Sternheim_Walker_1972}.

First-order perturbation theory consists of projecting the perturbation $V$ onto the ground state manifold
\begin{equation}\label{eq:1storderpert}
    \left\langle \psi_{1}^{L}\left|V\right|\psi_{1}^{R}\right\rangle.
\end{equation}
The wavefunction of the SCP is a combination of $M=0$ singlet and triplet components [see Eq.~\eqref{eq:states_2spins}], and, as such, the projection of $V$ vanishes. This is because, first, the operators $S_{2,3}^{\pm}$ do not conserve the value of $M$. Second, a simple direct calculation shows that $\left\langle \psi_{1}^{L}\left|S_{2,3}^{z}\right|\psi_{1}^{R}\right\rangle =0$.

To second order in perturbation theory, the correction to the Hamiltonian reads
\begin{equation} \label{eq:sec_order_PT_inter}
\Delta H^{(2)}=\sum_{j=2}^{4}\frac{\langle\psi_{1}^{L}|V|\psi_{i}^{R}\rangle\langle\psi_{i}^{L}|V|\psi_{1}^{R}\rangle}{-G_{j\to1}}.
\end{equation}
Here, we are not explicitly including the states of sites $1$ and $4$. The effective Hamiltonian $\Delta H^{(2)}$ will act on those sites, and the main task is to find the effective couplings. Neglecting constant terms, by simply evaluating Eq.~\eqref{eq:sec_order_PT_inter}, we find
\begin{equation}\label{eq:sec_order_PT}
   \Delta H^{(2)}=\tilde{J}_{1}\Bigg[\frac{e^{\tilde{\gamma}_{1}}}{2}S_{1}^{+}S_{4}^{-}+\frac{e^{-\tilde{\gamma}_{1}}}{2}S_{1}^{-}S_{4}^{+}+\tilde{\Delta}_{1}S_{1}^{z}S_{4}^{z}\Bigg].
\end{equation}
where
\begin{equation}\label{eq:sdrg_Herm_rule}
    \Tilde{J}_{1} = \frac{J_{1} J_{3}}{J_2} \frac{1}{1+\Delta_2}, \hspace{0.3cm} \Tilde{\Delta}_{1} = \frac{\Delta_{1} \Delta_{3}(1+\Delta_{2})}{2},
\end{equation}
\begin{equation}\label{eq:sdrg_nonHerm_rule}
    \Tilde{\gamma}_{1} = \gamma_{1} + \gamma_2 + \gamma_{3}.
\end{equation}
The $\mathrm{SO}(2)$ and PT symmetries assure that the form of the Hamiltonian under the SDRG decimations remains the same, as the RG does not break symmetries, and the starting Hamiltonian already is the most general respecting those symmetries.
The Hamiltonian $\Delta H^{(2)}$ has, therefore, the same form as Eq.~\eqref{eq:two-site}. Sites 1 and 4 are now connected with  effective couplings $\Tilde{J}_{1}, \Tilde{\gamma}_{1}, \Tilde{\Delta}_{1}$ given by Eqs.~\eqref{eq:sdrg_Herm_rule} and \eqref{eq:sdrg_nonHerm_rule}. 

This procedure is then iterated for the next largest gap. Notice that the gap between the ground state and first excited state is the same as the one for the Hermitian XXZ model so that the decimation in the nH case follows the same hierarchy of decimations as the Hermitian one would. The nH character is encoded both in the flow of $\gamma_i$ and in the ground-state wave function.

\section{Joint distributions of $J$, $\gamma$ and $\ell$ \label{sec:Distr}}
In order to calculate the physical properties of the chain at low energies,  it is necessary to calculate the joint fixed-point distribution of $J$, $\gamma$ and $\ell$, particularly at the cutoff $J=\Omega$, which we now proceed to do. Our discussion will focus on the XX and Heisenberg fixed-points, at which $\Delta_i=0$ and $\Delta_i=1$, respectively, and this variable does not appear in the distributions. We first change variables to  scaled ones, $\eta$, $x$ and $y$, via 
\begin{align}
\eta & =\frac{\ln\left(\Omega/J\right)}{\Gamma},\label{eq:eta_def}\\
x & =C_{\ell}\frac{\ell}{\Gamma^{2}},\label{eq:x_def}\\
y & =C_{\gamma}\frac{\gamma}{\Gamma^{n}} \label{eq:y_def}.
\end{align}
where $n=1$ for distributions that are even functions of $\gamma$, which we call symmetric, and $n=2$ when that is not the case, which we call asymmetric. The constants $C_\ell$ and $C_\gamma$ are non-universal and depend on the initial distributions.
Note that, although $\gamma$ and $\ell$ have the same decimation rules, see Eq.~(4) of the main text, lengths are always positive and asymmetrically distributed and, therefore, scaled by $\Gamma^2$. As in the main text, $\Gamma=\ln\left(\Omega_{0}/\Omega\right)$.
We call $P_{\Omega}\left(J,\ell,\gamma\right)$ the joint distribution of $J$, $\ell$ and $\gamma$, while $Q_{n}\left(\eta,x,y\right)$ is the distribution of scaled variables. At the fixed points, $Q_{n}\left(\eta,x,y\right)$ does not depend on $\Gamma$~\cite{fisher94-xxz}. From probability conservation,
\begin{equation}
P_{\Omega}\left(J,\ell,\gamma\right)dJd\ell d\gamma=Q_{n}\left(\eta,x,y\right)d\eta dx dy.\label{eq:pvsq}
\end{equation}
For couplings at the cutoff, $J=\Omega \Rightarrow \eta=0$.

Before discussing the joint distribution $Q_{n}\left(\eta,x,y\right)$, let us first summarize the results for the joint distribution obtained by marginalizing over $x$ (lengths), \cite{fisher94-xxz}
\begin{equation}
\label{eq:margin}
    q_{n}\left(\eta,y\right) = \int_{0}^{\infty}  Q_{n}\left(\eta,x,y\right) dx.
\end{equation}
As thoroughly discussed in reference~\cite{fisher94-xxz}, $q_{n}\left(\eta,y\right)$  can be, in general, reduced to a quadrature. Analytic expressions can be obtained~\cite{fisher94-xxz} for the distributions of $y$ marginalized over $\eta$
\begin{equation}
\label{eq:margin2}\overline{q}_{n}\left(y\right)=\int_0^{\infty} q_{n}\left(\eta,y\right)  d\eta,
\end{equation}
which are given by,
\begin{eqnarray}
\label{eq:q1bar}
\overline{q}_{1}\left(y\right)&=&\frac{1}{2\cosh\left(\pi y/2\right)},\\ \nonumber
\overline{q}_{2+}\left(y\right)&=&2\pi\sum_{n=0}^{\infty}\left(-1\right)^{n}\left(n+\frac{1}{2}\right)e^{-\left(n+\frac{1}{2}\right)^{2}\pi^{2}y}\\
\label{eq:q2barp}
&=&\frac{\pi}{2}\left.\frac{\partial\vartheta_1\left(z,e^{-\pi^2 y}\right)}{\partial z}\right|_{z=0},
\\
\label{eq:q2barm}
\overline{q}_{2-}\left(y\right)&=&\overline{q}_{2+}\left(-y\right),
\end{eqnarray}
where $\vartheta_{1}\left(z,q\right)$ is Jacobi's 1st theta function~\cite{abramowitzstegun1972} and $\overline{q}_{2+}\left(y\right)$ only has support for $y>0$, while $\overline{q}_{2-}\left(y\right)$ only has support for $y<0$. Although not explicitly given in ref.~\cite{fisher94-xxz}, it is easy to obtain analytic expressions for the distributions at the cutoff $q_{n}\left(\eta=0,y\right)$, which are~\cite{Hoyos2007PRB}
\begin{eqnarray}
\label{eq:q1}
q_{1}\left(\eta=0,y\right)&=&\frac{\pi/4}{\cosh^2\left(\pi y/2\right)},\\ \nonumber
q_{2+}\left(\eta=0,y\right)&=&-2\pi^{2}\sum_{n=1}^{\infty}\left(-1\right)^{n}n^{2}e^{-n^{2}\pi^{2}y}\\
\label{eq:q2p}
&=&-\pi^{2}e^{-\pi^{2}y}\left.\frac{\partial\vartheta_{4}\left(0,q\right)}{\partial q}\right|_{q=e^{-\pi^{2}y}},\\
\label{eq:q2m}
q_{2-}\left(\eta=0,y\right)&=&q_{2+}\left(\eta=0,-y\right),
\end{eqnarray}
where $\vartheta_{4}\left(z,q\right)$ is Jacobi's 4th theta function~\cite{abramowitzstegun1972} and analogous conditions hold for the support of $q_{2\pm}\left(\eta=0,y\right)$.

In order to obtain the full distributions  $Q_{n}\left(\eta,x,y\right)$, one must generalize the methods of ref.~\cite{fisher94-xxz}. But again, they can be reduced to a quadrature. The derivation is lengthy and details will be provided elsewhere. For our purposes here, we will only need their behavior at the cutoff $\eta=0$. For the case in which $\gamma$ is symmetrically distributed,
\begin{equation}
Q_{1}\left(\eta=0,x,y\right)=\frac{e^{-y^{2}/4x}}{2\sqrt{\pi x}}q_{2+}\left(\eta=0,x\right),
\label{eq:Q_xy_a-s}
\end{equation}
while for the case in which $\gamma$ is asymmetrically distributed,

\begin{equation}
Q_{2,\pm}\left(\eta=0,x,y\right)=\delta\left(x\mp y\right)q_{2+}\left(\eta=0, x\right),
\label{eq:Q_xy_a-a}
\end{equation}
where the upper (lower) sign refers to the case in which $y$, and hence $\gamma$, is positive(negative)-definite. 
It is immediate to see that Eq.~\eqref{eq:Q_xy_a-a}, when integrated over $x$, yields the expected result $q_{2\pm}\left(\eta=0,y\right)$.
The fact that marginalizing $Q_{1}\left(\eta=0,x,y\right)$ over $x$ leads to $q_{1}\left(\eta=0,y\right)$ is less obvious. But plugging the series of Eq.~\eqref{eq:q2p} into Eq.~\eqref{eq:Q_xy_a-s} and integrating over $x$ yields a series representation of the function in Eq.~\eqref{eq:q1}.

\begin{figure*}
    \centering
    \includegraphics[width=\linewidth]{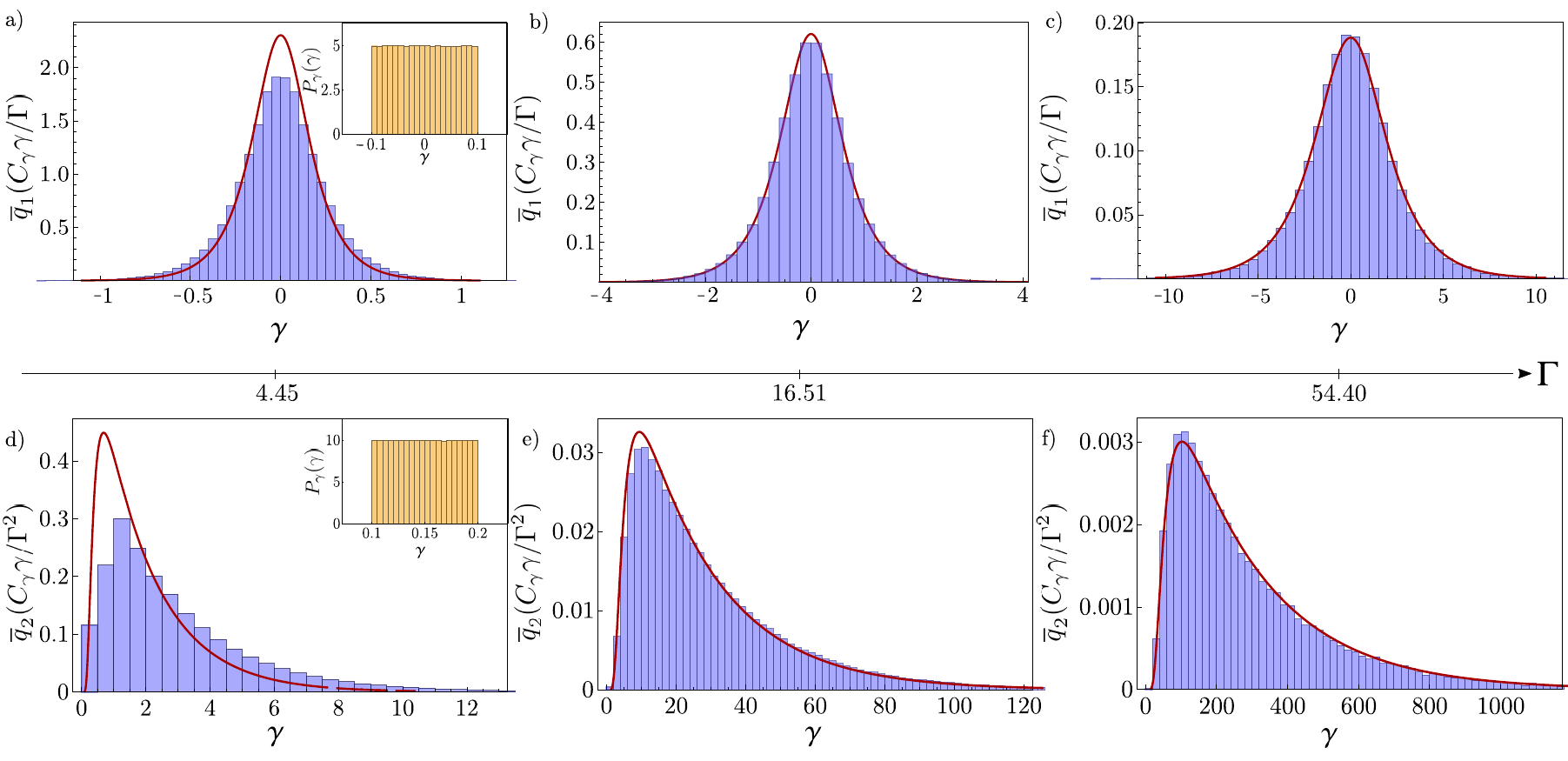}
    \caption{Three representative stages of the SDRG flow of the marginal distribution of the coupling $\gamma$, as $\Gamma$ increases. The histograms were obtained numerically for 25 disorder realizations of chains with 5 million sites 
    with periodic boundary conditions. The panels (a)-(c) correspond to an initial distributions that is symmetric, chosen as uniform with $-0.1 \leq \gamma \leq 0.1$ [see the inset of (a)]. The panels (d)-(f) correspond to a uniform asymmetric  initial distribution, in the range $0.1 \leq \gamma \leq 0.2$ [as shown in the inset of (d)]. The curves in red are the fixed point distributions (large $\Gamma$) from Eqs.~\eqref{eq:q1bar} [(a)-(c)] and~\eqref{eq:q2barp} [(d)-(f)]. By fitting to the distributions from the late stages of the SDRG flow [(c) and (f)], we determined the non-universal constants for (c) $C_{\gamma} = 20.5$ and (f) $C_{\gamma} = 4.8$.}
    \label{fig:marginal-distributions}
\end{figure*}

We have numerically simulated the SDRG flow for chains of 5 million sites averaging over 25 disorder realizations.  We could then fit the numerical histograms at late flow stages to the fixed-point distributions of Eqs.~\eqref{eq:q1bar} and~\eqref{eq:q2barp}, and thus determine the constant $C_{\gamma}$, defined in Eq.~\eqref{eq:y_def}, for both the symmetric and asymmetric cases, as can be seen in Fig.~\ref{fig:marginal-distributions}. This constant is non-universal and depends on the bare initial distributions. We have used initial uniform distributions, as shown in yellow in the insets. The agreement is remarkable.

\section{The magnetic susceptibility due to the strongly coupled pairs\label{sec:magn_suscep}}

In this Section we provide details on how to calculate the contribution to the magnetic susceptibility due to the SCPs. We first calculate the susceptibility of a single SCP. This is then used to collect the contribution from all the SCPs formed along the SDRG flow.

\subsection{The magnetic susceptibility of a single SCP}

We will calculate the magnetic susceptibility of a single pair by adding a small magnetic field term to the SCP Hamiltonian and taking the zero-field limit
\begin{equation}
    H = H_0^{\text{NH}} - \textbf{h}\cdot \textbf{S},
\end{equation}
where $H_0^{\text{NH}}$  is the nH Hamiltonian of one SCP without the magnetic field. The canonical partition function reads
\begin{equation}
    Z = \text{Tr} (e^{-\beta H}) = \sum_n \langle \psi_n |e^{-\beta H}| \psi_n\rangle,
\end{equation}
with $\beta=1/T$ (we take $k_{B}=1$).
As usual, the trace can be taken with any complete basis. Since we are working with a nH system, we face the possible ambiguity of calculating the expectation values with either the left-right eigenvectors or the right-right (or, equivalently, left-left) ones~\cite{Brody_2014, fossati_entanglement_2023}. Notice, however, that only the left-right ones are complete. Therefore, these are the ones that we use in our derivation. We show next that, with this choice, the computation follows the same steps as in a Hermitian model, as the same thermodynamic identities hold.

The free energy is given by $F=-(\ln Z)/\beta$, while the $\alpha$ component  of the magnetization reads
\begin{equation}
\begin{split}
    M^{\alpha} &= - \frac{\partial F}{\partial h^{\alpha}}\Bigg|_{h^{\alpha} \to 0},\\ &=\frac{1}{\beta\text{Tr}e^{-\beta H}}\sum_{n}\left\langle \psi_{n}^L\left|\frac{\partial \left[e^{-\beta H}\right]}{\partial h^{\alpha}}\right|\psi_{n}^R\right\rangle \Bigg|_{h^{\alpha}\to0},\\
    &=\langle S^{\alpha} \rangle,
\end{split}    
\end{equation}
where $S^{\alpha}$ is the total angular momentum along $\alpha$. The diagonal components $\chi^{\alpha}_{\mathrm{sscp}}$ of the magnetic susceptibility of a single SCP follow from $M^{\alpha}$,
\begin{equation}
    \chi^{\alpha}_{\mathrm{sscp}} = \frac{\partial M^{\alpha}}{\partial h^{\alpha}}\Bigg|_{h^{\alpha} \to 0}=\frac{1}{T}\left[\left\langle \left(S^{\alpha}\right)^{2}\right\rangle -\left\langle S^{\alpha}\right\rangle^2 \right].
\end{equation}

For our particular Hamiltonian, the two-site ground state is a linear combination of the singlet and the $M=0$ triplet, so $\left\langle S^{\alpha}\right\rangle =0$ and 
\begin{equation}
     \chi^{\alpha}_{\mathrm{sscp}} =\frac{1}{T} \langle \left(S^{\alpha}\right)^{2} \rangle.
\end{equation} 
We calculate the mean value in the ground state, ignoring thermal fluctuations to excited states at low $T$, so $\langle \left(S^{\alpha}\right)^{2} \rangle = \langle \psi^L_1 | \left(S^{\alpha}\right)^{2} | \psi^R_1\rangle$. This state is given in Eq.~\eqref{eq:states_2spins}. The mean value vanishes for $\alpha=z$ and
\begin{equation}
\langle \left(S^{\alpha}\right)^{2} \rangle =
\begin{cases}
-\sinh^{2}\left(\frac{\gamma}{2}\right), & \ \text{if } \alpha = x,y, \\
0, & \ \text{if } \alpha = z,
\end{cases}
\end{equation}
so that a single SCP  only contributes  to the magnetic susceptibility in the $x-$ and $y-$directions 
\begin{equation}
\chi^{x,y}_{\mathrm{sscp}}  = -\frac{\sinh^{2}\left(\frac{\gamma}{2}\right)}{T}.\label{eq:chixysscp}
\end{equation}
This is Eq.~(8) of the main text.

\subsection{The magnetic susceptibility due to all the SCPs}

\begin{figure}
    \centering
    \includegraphics[width=\linewidth]{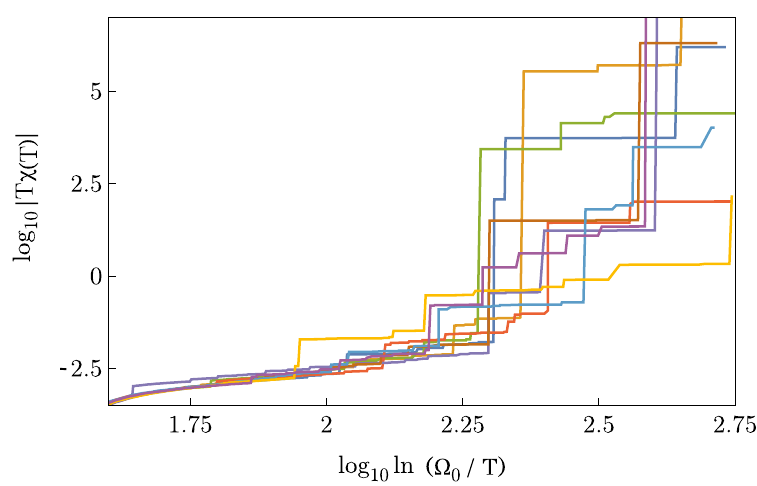}
    \caption{The magnetic susceptibility for nine different realizations of the same initial uniform distribution, $-0.1 \leq \gamma \leq 0.1$, with periodic boundary conditions. There are huge sample to sample fluctuations and jumps of all sizes at different temperatures. This illustrates why the average over realizations shown in Fig.~2 of the main text fluctuates even for several disorder realizations.}
    \label{fig:divergence_susc}
\end{figure}

We now have to sum over all the SCPs that were decimated from the initial scale $\Omega=\Omega_0$ down to the scale $\Omega=T$. At an intermediate scale $\Omega$, there is a density of $n_{\Omega}$ undecimated spins, of which $(n_{\Omega}/2)P_{\Omega}(J=\Omega,\gamma)d\Omega$ will form SCPs in the next sliver $d\Omega$ of decimations. Note that the bond lengths are immaterial here, so we have omitted $\ell$. Summing their contributions to the $x,y$ susceptibilities, Eq.~\eqref{eq:chixysscp}, from $\Omega_0$ to $T$,
\begin{eqnarray}
        \chi^{x,y}_{\text{SCP}}\left(T\right) & =&-\frac{1}{T}\int_{T}^{\Omega_{0}}d\Omega\frac{n_\Omega}{2}\int d\gamma \sinh^{2}\left(\frac{\gamma}{2}\right) \nonumber \\ && \times P_{\Omega}(J=\Omega,\gamma),
\end{eqnarray}
where the $\gamma$ integration runs over the support of the distribution, whatever it may be. We are interested in the low temperature behavior, for which we can use the fixed-point distributions.
Changing to the variables in Eqs.~\eqref{eq:eta_def} and~\eqref{eq:y_def} and to $\Gamma=\ln(\Omega_0/\Omega)$, using Eq.~\eqref{eq:pvsq}, and the definition in Eq.~\eqref{eq:margin},
we obtain
\begin{eqnarray}
\chi^{x,y}_{\text{SCP}}\left(T\right) &=&-\frac{1}{2T}\int_{0}^{\Gamma_{T}}\frac{d\Gamma}{\Gamma}\overline{n}_{\Gamma}\nonumber \\ &&\times \int dy\sinh^{2}\left(\frac{y\Gamma^{n}}{2C_{\gamma}}\right)q_{n}(\eta=0,y),
\end{eqnarray}
where $\Gamma_T=\ln(\Omega_0/T)$ and $\overline{n}_{\Gamma}=n_{\Omega_0 e^{-\Gamma}}$.
The distributions $q_n\left(\eta=0,y\right)$ were given in Eqs.~\eqref{eq:q1}-\eqref{eq:q2m}.
Now, the integral over $y$ 
\begin{equation}
g_{n}\left(\Gamma\right)=\int dy\sinh^{2}\left(\frac{y\Gamma^{n}}{2C_{\gamma}}\right)q_{n}\left(\eta=0,y\right)
\end{equation}
diverges for large enough $\Gamma$ (low enough temperature). Defining this threshold value as $\Gamma^{(c)}$, we can find it from the asymptotic behavior of $q_n\left(\eta=0,y\right)$ for large $y$
\begin{eqnarray}
    q_1\left(\eta=0,y\right) &\sim& e^{-\pi |y|},\\
    q_{2\pm}\left(\eta=0,y\right) &\sim& e^{-\pi^2 |y|}.
\end{eqnarray}
For the symmetric case,
\begin{equation}
\frac{y\Gamma_{\text{sym}}^{\left(c\right)}}{C_{\gamma}}-\pi y=0\implies\Gamma_{\text{sym}}^{\left(c\right)}=\pi C_{\gamma},
\end{equation}
whereas for the asymmetric case
\begin{equation}
\frac{y\left(\Gamma_{\text{asym}}^{\left(c\right)}\right)^{2}}{C_{\gamma}}-\pi^{2}y=0\implies\Gamma_{\text{asym}}^{\left(c\right)}=\pi\sqrt{C_{\gamma}}.
\end{equation}
In the symmetric case, calling $A=\frac{\Gamma}{2C_{\gamma}}$,
the integral can be explicitly computed,
\begin{eqnarray}
  g_n(\Gamma)&=&\int_{-\infty}^{\infty}dy\sinh^{2}\left(Ay\right)q_1\left(\eta=0,y\right)\nonumber \\
  &=&\frac{1}{2}\left(\frac{2A}{\sin\left(2A\right)}-1\right),  
\end{eqnarray}
as long as $A<\pi/2$. Using the asymptotic behavior $\overline{n}_{\Gamma} \sim \Gamma^{-2}$~\cite{fisher94-xxz}, we get
\begin{align}
\chi_{\text{SCP}}\left(T\right) & =-\frac{1}{4T}\ln\left[\frac{2C_{\gamma}}{\Gamma_{T}}\tan\left(\frac{\Gamma_{T}}{2C_{\gamma}}\right)\right].
\end{align}
This is Eq.~(9) of the main text.
It confirms a log-divergence
at $T=\Omega_{0}\exp\left(-\pi C_{\gamma}\right)$. A divergence is found numerically as well, as seen in Fig.~2 of the main text, but the value of $T$ at which it occurs is hard to pinpoint precisely. As shown in Fig.~\ref{fig:divergence_susc}, each realization can have more than one sudden enhancement in the value of $T\chi$ and these discontinuities happen at different values of $T$ for each realization. In other words, there are large sample to sample fluctuations of the susceptibility, which is not self-averaging.

The large fluctuations of $\chi_{\text{SCP}}$ can be understood from the distribution, \textit{for a fixed} $\Gamma$, of
\begin{align}
V & =-\sinh^{2}\left(\gamma/2\right)\equiv T\chi^{x,y}_{sscp},\\
 & \sim-\frac{1}{4}e^{|\gamma|},\,\,|\gamma|\gg1.
\end{align}
We will call its distribution $\mathcal{V}_{n}\left(V\right)$, with $n=1,2$.
For the symmetric case, from Eq.~\eqref{eq:q1},
\begin{align}
q_{1}\left(\eta=0,y=\frac{C_{\gamma}\gamma}{\Gamma}\right) & =\frac{\pi/4}{\cosh^{2}\left(\frac{\pi C_{\gamma}\gamma}{2\Gamma}\right)},\\
 & \sim \pi e^{-\frac{\pi C_{\gamma}|\gamma|}{\Gamma}},\,\,\frac{\pi C_{\gamma}|\gamma|}{\Gamma}\gg1,
\end{align}
and the distribution $\mathcal{V}_{\mathrm{1}}\left(V\right)$ has the asymptotic behavior for large $V$
\begin{align}
\mathcal{V}_{\mathrm{1}}\left(V\right) & =\left|\frac{d\gamma}{dV}\right|\frac{\pi/4}{\cosh^{2}\left(\frac{\pi C_{\gamma}\gamma}{2\Gamma}\right)},\\
 & \sim\left|\frac{1}{V}\right|^{1+\frac{\pi C_{\gamma}}{\Gamma}}.
\end{align}
It is clear that $\langle V\rangle$ diverges when $\Gamma \ge \pi C_\gamma=\Gamma^{(c)}_{\mathrm{sym}}$, as we saw. However, $\langle V^2\rangle$ diverges even before that, when $\Gamma \ge \pi C_\gamma/2=\Gamma^{(c)}_{\mathrm{sym}}/2$. This explains the large fluctuations that affect the numerically determined average susceptibility.

For the asymmetric case, from Eqs.~\eqref{eq:q2p} and~\eqref{eq:q2m},
\begin{align}
q_{2\pm}\left(0,y=\frac{C_{\ell}\gamma}{\Gamma^{2}}\right) & =-2\pi^{2}\sum_{k=1}^{\infty}\left(-1\right)^{k}k^{2}e^{-k^{2}\pi^{2}C_{\gamma}|\gamma|/\Gamma^{2}},\\
 & \sim2\pi^{2}e^{-\pi^{2}C_{\gamma}|\gamma|/\Gamma^{2}},\,\,\frac{\pi^2 C_{\gamma}|\gamma|}{\Gamma^2}\gg1.
\end{align}
The distribution is, again, a power law
\begin{equation}
\mathcal{V}_{\mathrm{2}}\left(V\right)\sim\frac{1}{\left|V\right|^{1+\pi^{2}C_{\gamma}/\Gamma^{2}}},
\end{equation}
and $\langle V\rangle$ diverges at $\Gamma=\Gamma_{\text{asym}}^{\left(c\right)}$, but $\langle V^2\rangle$ diverges before that, when $\Gamma \ge \pi \sqrt{C_\gamma/2}=\Gamma^{(c)}_{\mathrm{asym}}/\sqrt{2}$, again explaining the large fluctuations observed numerically.

\section{The entanglement entropy (EE) \label{sec:EE}}
In this Section, we derive the entanglement entropy of a partition of size $L$ with the rest of the infinite chain using the SDRG framework. We will follow the arguments of Ref.~\cite{Hoyos2007PRB}, with the proper adjustments needed to take into account that the strongly coupled pairs (SCPs) are formed by a mixture of singlet and triplet components. As was the case in the calculation of the magnetic susceptibility, the EE of the partition also requires that we first calculate the EE between spins in an SCP.
\subsection{EE in a single SCP}
We start by computing the EE between spins in a single SCP in its ground state, $S_{\text{pair}}$. Non-Hermiticity implies that there are two ways of defining the ground state density matrix of the SCP \cite{couvreur_entanglement_2017,Herviou_entanglement_2019,guo_entanglement_2021,modak_entanglement_2021,fossati_entanglement_2023,Cipolloni_entanglement_2023,Hsieh_entanglement_2023,kawabata_entanglement_2023}. These ways differ in the way you normalize the ground state. We will call them RR and RL schemes. 

In the RR scheme, the ground state of an SCP is normalized as $\langle\psi_{\text{GS}}^{\text{RR}}|\psi_{\text{GS}}^{\text{RR}}\rangle=1$ and reads
\begin{align}
\left|\psi_{\text{GS}}^{\text{RR}}\right\rangle  & =\frac{1}{\sqrt{\cosh\gamma}}\left(\cosh\left(\frac{\gamma}{2}\right)\left|0,0\right\rangle +\sinh\left(\frac{\gamma}{2}\right)\left|1,0\right\rangle \right),\nonumber \\
 & =\frac{1}{\sqrt{2\cosh\gamma}}\left(e^{\gamma/2}\left|+-\right\rangle +e^{-\gamma/2}\left|-+\right\rangle \right).
\end{align}
The density matrix is $\rho^{\text{RR}}=\left|\psi_{\text{GS}}^{\text{RR}}\right\rangle \left\langle \psi_{\text{GS}}^{\text{RR}}\right|$. The reduced density matrix is found by taking the partial trace with respect to any one of the sites, say site 2,
\begin{equation}
\begin{split}
    \rho^{\text{RR}}_{1} &= \text{Tr}_{2} \ \rho^{\text{RR}}, \\
    &=\frac{1}{2\cosh \gamma} \left(e^\gamma|+\rangle\langle +| + e^{-\gamma}|-\rangle\langle-|\right).
\end{split}
\end{equation}
The EE of either spin of the pair follows easily
\begin{equation}
\begin{split}
    S_{\text{pair}} &= -\text{Tr} \left(\rho_1^{\text{RR}}\log_2\rho_1^{\text{RR}}\right)\\
    &=|\gamma|(1-\tanh{|\gamma|})\log_2{e} + \log_2(1+e^{-2|\gamma|}),
    \label{eq:EE_SCP}
\end{split}
\end{equation}
which is Eq.~(11) of the main text. Note that $S_{\mathrm{pair}}=1$ in the nH limit and $S_{\mathrm{pair}}\to 0$
as $\gamma \to \infty$, which shows that extreme non-Hermiticity kills any quantum entanglement. Thus, the RR scheme captures efficiently the effects of non-Hermiticity. 
An LL scheme, in which $\langle\psi_{\text{GS}}^{\text{LL}}|\psi_{\text{GS}}^{\text{LL}}\rangle=1$, would involve simply changing $\gamma \to -\gamma$ above and is equivalent to the RR scheme.

In the RL scheme, the ground state is normalized according to $\langle \psi_{GS}^L| \psi_{GS}^R\rangle = 1$ and, as given in Eq.~\eqref{eq:states_2spins},
\begin{equation}
\begin{split}
    &|\psi^R_{GS}\rangle = \frac{1}{\sqrt{2}}\left(e^{-\gamma}|+-\rangle - |-+\rangle\right),\\
    &|\psi^L_{GS}\rangle = \frac{1}{\sqrt{2}}\left(e^{\gamma}|+-\rangle -
 |-+\rangle\right).
\end{split}
\end{equation}
The density matrix is then $\rho^{\text{RL}} = |\psi_{GS}^\text{R}\rangle\langle \psi_{GS}^\text{L}|$ with the reduced density matrix
\begin{equation}
    \rho_1^{RL} = \text{Tr}_2 \ \rho^{RL} = \frac{1}{2}\left(|+\rangle\langle +| + |- \rangle\langle -|\right).
\end{equation}
It follows that $S_{\text{pair}} = -\text{Tr} \left(\rho_1^{\text{RL}}\log_2\rho_1^{\text{RL}}\right) =1$, the same result as for a singlet. In the RL scheme, therefore, the EE does not distinguish between the Hermitian and nH cases.

\subsection{The EE in the RSCPP}

We now tackle the calculation of the EE of a partition of size $L$ with the rest of the chain fixed in its ground state, $S\left(L\right)$. The partition is entangled with the rest of the chain through the SCPs that cross either of its ends~\cite{RefaelMoore_PRL_2004, Refael_2009}. All one has to do is to sum up the contributions of all these end-crossing SCPs. Since the SCPs are formed at different intermediate scales $\Omega$, one has follow the SDRG flow from the initial scale $\Omega_0$ down to a low-energy scale $\Omega(L)$, such that for $\Omega<\Omega(L)$, only SCPs with bond lengths larger than $L$ are formed and, therefore, cannot connect spins inside the partition with spins outside it. As explained in Appendix C of ref.~\cite{Hoyos2007PRB}, this can be done by summing over the
lengths $\ell$ of all bonds decimated until the energy scale $\Omega(L)$ while weighting them with the EE of a SCP with nH parameter $\gamma$ as given in Eq.~\eqref{eq:EE_SCP}. Thus, 
\begin{equation}
\label{eq:seeL}
S\left(L\right)=2A_n\left[\Omega(L)\right].
\end{equation}
with 
\begin{align}
A_n\left(\Omega\right) & =\int_{\Omega}^{\Omega_{0}}d\Omega n_{\Omega}\int_0^{\infty} d\ell \int d\gamma P_{\Omega}\left(\Omega,\ell,\gamma\right)\ell S_{\text{pair}}\left(\gamma\right).\label{eq:M_integral}
\end{align} 
The factor of $2$ in Eq.~\eqref{eq:seeL} accounts for the contributions from both partition ends, the subscript $n=1,2$ distinguishes between the symmetric and asymmetric $\gamma$ distributions, whose supports define the limits of the $\gamma$ integration, as already discussed.
Changing to the variables in Eqs.~\eqref{eq:eta_def}-\eqref{eq:y_def} and to $\Gamma=\ln(\Omega_0/\Omega)$, and using the asymptotic behavior $\overline{n}_{\Gamma}\sim1/\ \Gamma^{2}$,
\begin{align}
A_{n}\left(\Omega\right) & =\frac{1}{C_{\ell}}\int_{\Gamma_{0}}^{\Gamma}\frac{d\Gamma}{\Gamma} \int dyS_{\text{pair}}\left(y\Gamma^{n}/C_{\gamma}\right)\times\nonumber \\
 & \times\int_{0}^{\infty}dx\,\,xQ_{n}\left(\eta=0,x,y\right).\label{eq:An0}
\end{align}
Note that we left the initial scale as $\Gamma_0\neq 0$ because the fixed-point behavior we are assuming is only valid for $\Omega \ll \Omega_0$. This arbitrariness will translate into a non-universal additive constant, as we will see. Now, $S_{\mathrm{pair}}(y)$ is an even function of $y$. In the symmetric case, $n=1$, $Q_{n}\left(\eta=0,x,y\right)$ is also even in $y$. In that case, we can set the limits of the $y$ integral from $0$ to $\infty$ and multiply the result by 2. In the asymmetric case, $n=2$, we can use Eq.~\eqref{eq:Q_xy_a-a} to see that (1) we can set the same $y$ integration limits and (2) both positive- and negative-definite distributions lead to the same result. Thus,
\begin{equation}
    A_n\left(\Omega\right)=\frac{2}{nC_{\ell}}\int_{\Gamma_{0}}^{\Gamma}\frac{d\Gamma}{\Gamma} \int_0^{\infty} dyS_{\text{pair}}\left(y\Gamma^{n}/C_{\gamma}\right) \Lambda_n(y),
\end{equation}
where we defined
\begin{equation}
\Lambda_n(y)=\int_{0}^{\infty}dx\,\,xQ_{n}\left(\eta=0,x,y\right).
\end{equation}
Defining $y= C_{\gamma}u/\Gamma^{n}$,
\begin{equation}
    A_{n}\left(\Omega\right) = \frac{2C_\gamma}{nC_\ell}
    \int_0^{\infty} S_{\text{pair}}\left(u\right)du\int_{\Gamma_{0}}^{\Gamma}\frac{d\Gamma}{\Gamma^{n+1}}  \Lambda_n\left( C_{\gamma}u/\Gamma^{n}\right),
\end{equation}
and changing variable from $\Gamma$ to $t=C_{\gamma}u/\Gamma^{n}$, we obtain
\begin{equation}
    A_{n}\left(\Omega\right) =
    \frac{2}{n^2C_{\ell}}\int_0^{\infty} du\frac{S_{\text{pair}}\left(u\right)}{u}\int_{t=C_{\gamma}u/\Gamma^{n}}^{t_{0}=C_{\gamma}u/\Gamma_{0}^{n}}
    \Lambda_n(t)dt.\label{eq:An}
\end{equation}
For the asymmetric case, $n=2$, using Eq.~\eqref{eq:Q_xy_a-a}, 
\begin{equation}
A_{2}\left(\Omega\right)=\frac{1}{2C_{\ell}}\int_0^{\infty} du\frac{S_{\text{pair}}\left(u\right)}{u}\int_{t}^{t_{0}}t q_{2+}\left(\eta=0,t\right)dt.
\end{equation}
Plugging the series of Eq.~\eqref{eq:q2p}, into the integral
\begin{eqnarray}
\int_0^tdt\,tq_{2+}\left(t\right)&=&2\sum_{n=1}^{\infty}\left(-1\right)^{n}\left[e^{-n^{2}\pi^{2}t}\left(t+\frac{1}{\pi^{2}n^{2}}\right)\right.\nonumber \\&-&\left.\frac{1}{\pi^{2}n^{2}}\right] \label{eq:definite}
\end{eqnarray}
allows us to write 
\begin{equation}
A_{2}\left(\Omega\right)=B_{2}\left(\Omega\right)-B_{2}\left(\Omega_{0}\right),\label{eq:A2}
\end{equation} where
\begin{widetext}    
\begin{eqnarray}
B_{2}\left(\Omega\right)=\frac{1}{C_{\ell}}\sum_{n=1}^{\infty}\left(-1\right)^{n}\int_{0}^{+\infty}du\frac{S_{\text{pair}}\left(u\right)}{u}\left[\frac{1}{\pi^{2}n^{2}}-e^{-n^{2}\pi^{2}C_{\gamma}u/\Gamma^{2}}\left(\frac{C_{\gamma}u}{\Gamma^{2}}+\frac{1}{\pi^{2}n^{2}}\right)\right]. \label{eq:b2}
\end{eqnarray}
\end{widetext}    
Note that the use of the definite integral in Eq.~\eqref{eq:definite}
guarantees the convergence of the integral in Eq.~\eqref{eq:b2} at its lower limit.
We see that, as $\Omega \to 0$ or $\Gamma \to \infty$, $B_{2}\left(\Omega \to 0 \right) \rightarrow 0$. This reflects the fact that long SCPs formed at low energies show no entanglement, as they tend to have $\gamma\gg 1$ [see Eq.~\eqref{eq:EE_SCP} and the discussion thereafter]. The strong non-Hermiticity renders them completely classical.

For the symmetric case, using Eq.~\eqref{eq:Q_xy_a-s},
\begin{equation}
\Lambda_1\left(t\right)=\int_{0}^{\infty}dx\,x \frac{e^{-t^{2}/4x}}{2\sqrt{\pi x}}q_{2+}\left(x\right).
\end{equation}
Using the series for $q_{2+}\left(x\right)$, Eq.~\eqref{eq:q2p}, integrating
term by term, and re-summing we obtain 
\begin{equation}
\Lambda_1\left(t\right)=\frac{1}{2\pi}\left[\frac{\pi\left|t\right|}{1+e^{\pi\left|t\right|}}+\ln\left(1+e^{-\pi\left|t\right|}\right)\right].
\end{equation}
The remaining integral over $t$ can also be calculated,
\begin{eqnarray}
\int dt\,\Lambda_1\left(t\right) & = & \frac{\mathrm{Li}_{2}\left(-e^{-\pi t}\right)}{\pi^2}-\frac{t\ln\left(1+e^{-\pi t}\right)}{2\pi},\label{eq:Lambda1_int}
\end{eqnarray}
where $\mathrm{Li}_{2}(x)$ is the polylogarithm function of order 2, defined
by
\begin{align}
\mathrm{Li}_{2}\left(z\right) & =-\int_{0}^{z}dt\frac{\ln\left(1-t\right)}{t},\\
 & =\sum_{k=1}^{\infty}\frac{z^{k}}{k^{2}}.
\end{align}
Plugging the result of Eq.~\eqref{eq:Lambda1_int} into Eq.~\eqref{eq:An} with $n=1$, we can write 
\begin{equation}
A_{1}\left(\Omega\right)=B_{1}\left(\Omega\right)-B_{1}\left(\Omega_{0}\right), \label{eq:A1}
\end{equation}
where
\begin{widetext}    
\begin{eqnarray}
B_{1}\left(\Omega\right)=\frac{1}{\pi C_{\ell}}\int_{0}^{\infty}du\frac{S_{\text{pair}}\left(u\right)}{u}\left\{ \frac{C_{\gamma}u}{\Gamma}\ln\left(1+e^{-\pi C_{\gamma}u/\Gamma}\right)-\frac{2}{\pi}\left[\mathrm{Li}_{2}\left(-e^{-\pi C_{\gamma}u/\Gamma}\right)+\frac{\pi^{2}}{12}\right]\right\},\label{eq:B1}
\end{eqnarray}
\end{widetext}
and the term in $\pi^2/12$ was added to each term on the right-hand side of 
Eq.~\eqref{eq:A1}, so that the integral in Eq.~\eqref{eq:B1} converges at its lower limit.

Again, as $\Omega \to 0$ or $\Gamma \to \infty$, $B_{1}\left(\Omega \to 0 \right) \rightarrow 0$, showing that long SCPs formed late in the flow will no longer contribute to the EE in this limit. 

\begin{figure}
    \centering
    \includegraphics[width=\linewidth]{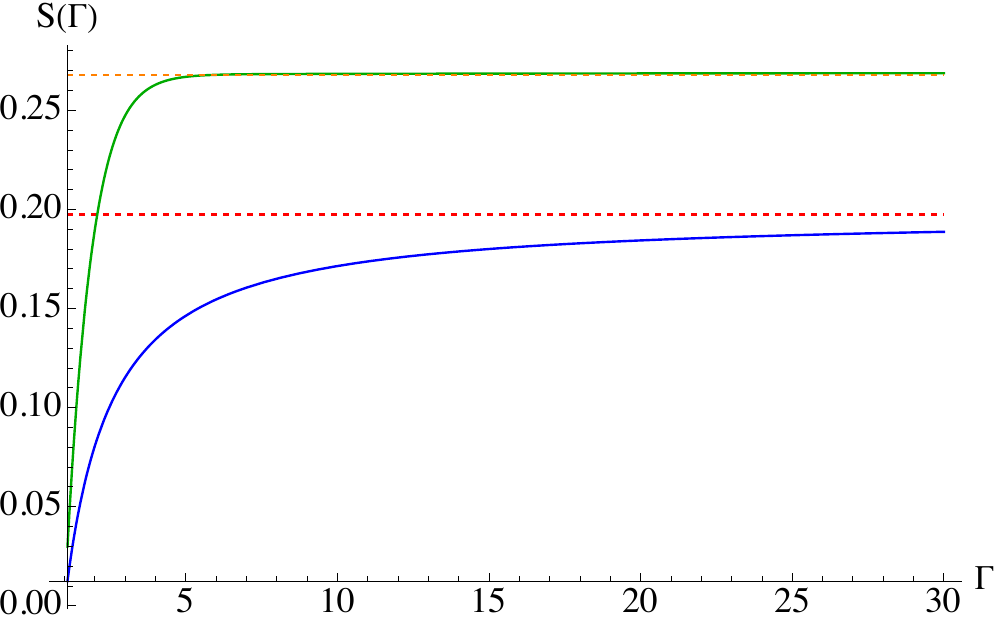}
    \caption{Entanglement entropy
    for symmetric (blue) and asymmetric (green) initial $\gamma$ distributions as a function of $\Gamma$. We chose $C_{\ell}=C_{\gamma}=1$. Other choices of $C_{\ell}$ and $C_\gamma$ correspond to a rescaling of the vertical and horizontal axes. The horizontal lines are the asymptotic values at which the entanglement entropy saturates. This saturation is in sharp contrast with the EE in the Hermitian case, which increases linearly with $\Gamma$. It reflects the destruction of quantum entanglement by non-Hermiticity.}
    \label{fig:EE_analytical}
\end{figure}

Now both results of Eqs.~\eqref{eq:A2} and~\eqref{eq:A1} can be used in Eq.~\eqref{eq:seeL}, together with the asymptotic form~\cite{fisher94-xxz}
\begin{equation}
    \Gamma \sim \sqrt{L} \Rightarrow
    \Omega\sim \Omega_0 e^{-\sqrt{L}},
\end{equation}
to obtain the desired EE as a function of partition size $L$. We show the EE as a function of $\Gamma=\ln(\Omega_0/\Omega)$ for both the symmetric and asymmetric cases in Fig.~\ref{fig:EE_analytical}. 
Note that the use of an arbitrary $\Omega_0$, and hence $\Gamma_0$, in Eqs.~\eqref{eq:A2} and~\eqref{eq:A1} amounts to a non-universal additive constant to $S(L)$ resulting from the initial flow. Besides, the other two non-universal constants, $C_\ell$ and $C_\gamma$, that also depend on the initial distribution, will determine the overall scale of the vertical and horizontal axes, respectively.

\end{document}